\documentclass{article}
\usepackage{bm,amsmath}
\usepackage{epsfig}
\date{\small Received April 4, 2000; in final form, September 13, 2000}
\author{{\bf V. N. Shalyapin
\footnote{E-mail address for contacts: vshal@ira.kharkov.ua}}\\
{\it \small Institute of Radio Astronomy, National Academy of
Sciences of Ukraine,}\\ {\it \small Kharkov, Ukraine}}
\title{\bf {Caustic Crossing in the Gravitational Lens Q2237+0305}}

\begin{document}
\maketitle
\section*{}
{\small The monitoring of the gravitational lens Q2237+0305
carried out by the OGLE group during 1997--1999 is analyzed. The
significant light amplifications in the C and A quasar components
with maxima in mid- and late 1999, respectively, are interpreted
as the crossing of microlens caustics by the source. A constraint
on the emitting-region size $R \leq 10^{15}$ cm has been obtained
from the light-curve shape by assuming a power-law quasar
brightness distribution $(r^2+R^2)^{-p}$. To estimate the exponent
$p \sim 1.2$ requires refining the standard model for the quasar
continuum formation in an optically thick accretion disk with $p =
1.5$.}

\section*{}
{{\bf Key words:}  quasars, gravitational lenses, accretion disks}

\section{INTRODUCTION}
Quasar variability under the effect of microlenses depends both on
parameters of the mass distribution for compact bodies and on the
appearance of the emitting region. For instance, the larger is the
quasar, the smaller are the amplitudes of light variations in its
images. Thus, it becomes possible to formulate the inverse problem
of determining the sizes and structure of quasars from their
observed light curves. The main difficulty in solving this problem
is that the distribution of the amplification produced by
microlenses is not known in advance and exhibits a fairly complex
pattern with many randomly located caustic lines (see, e.g.,
Zakharov 1997 and Zakharov and Sazhin 1998). In general, the
specific form of this distribution is not known and can be
analyzed only statistically. Exceptions are only those portions of
the light curve that correspond to caustic crossing by the quasar.
The amplification of a point source during caustic crossing obeys
a simple law: it remains approximately constant as the caustic is
approached, then abruptly increases to infinity at the caustic
itself, and subsequently falls off as $x^{-1/2}$ with increasing
distance $x$ from the caustic (Chang and Refsdal 1984; Blandford
and Narayan 1986).

Microlenses must show up most clearly in the multiple quasar
images produced by the gravitational effect of galaxy macrolenses.
First, the microlensing probability is rather high in such
situations. Second, intrinsic quasar variability can, in
principle, be separated from microlensing variability. Of all the
gravitationally lensed quasars, Q2237+0305 ($z_s=1.675$) is
undoubtedly the most promising object for microlensing analysis.
Because the unique proximity of a lensing galaxy ($z_l = 0.039$),
microlensing variability in this object must take place faster
than in other objects by an order of magnitude and with a large
amplitude.

The quasar Q2237+0305 has been monitored virtually since its
discovery, and it actually proved to be the first object in which
microlensing variability was detected (Irwin et al. 1989; Corrigan
et al. 1991). The observations by \O stensen et al. (1996) showed
that virtually all four quasar images were more or less variable.

The regularity and quality of the Q2237+0305 monitoring have
improved markedly when the OGLE group joined it in the last four
years (Wozniak et al. 2000) (see Fig. 1). Measurements are made in
the V band approximately once a week during the observing season
from May through December. The latest observations
(http://www.astro.princeton.edu/$\sim$ogle/ogle2/huchra.html) show
that image C passed its intensity peak in mid-1999, while image A
peaked in late 1999. Interpreting the light-curve maxima as
resulting from caustic crossing allows the size and structure of
the quasar emitting region in the object under study to be
determined.
\begin{figure}
\epsfig{file=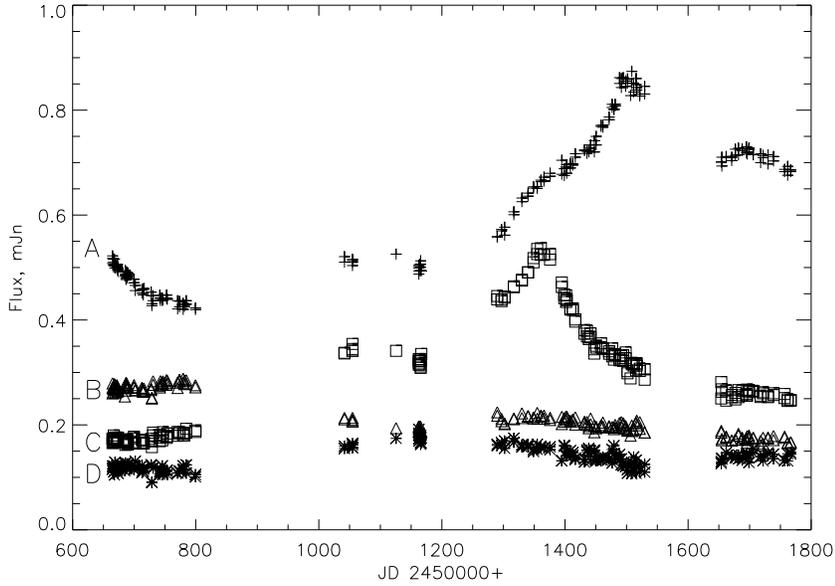,width=11cm}
\caption{V light variations of
the four images of the quasar Q2237+0305 during 1997--2000 as
observed by the OGLE group}
\end{figure}

\section{AMPLIFICATION OF AN EXTENDED\\ SOURCE IN THE CAUSTIC REGION}
When a source crosses a caustic line, an additional pair of images
appears (or disappears). The total intensity of this pair depends
on the distance to the caustic as $x^{-1/2}$. Therefore, the
intensity of a point source in the caustic region can be
represented as (Schneider and Weiss 1987)
\begin{equation}
I_p(x) = I_0 +\theta(x) a_0 x^{-1/2}.
\end{equation}
Here, $I_0$ is the intensity of all the remaining microimages
except the additional pair, $\theta(x)$ is the Heaviside unit
function, and $a_0$ is the caustic strength. The amplification of
an extended source with a brightness distribution $P(\bm{r})$ can
be calculated by ordinary summation over the set of infinitely
small sources with individual amplification factors:
\begin{equation}
I = \left[\int d\bm{r} P(\bm{r}) I_p(\bm{r})\right] \Big/
\left[\int d\bm{r} P(\bm{r}) \right].
\end{equation}
To describe the brightness distribution in the source, we use a
power-law model
\begin{equation}   \label{eq:P}
P(r) = (1+\frac{r^2}{R^2})^{-p} \qquad    (p>0),
\end{equation}
which is determined by the source radius $R$ and by the rate of
brightness decline $p$.

Let the source center be at a distance $D$ from the caustic line.
In the normalized coordinates $\xi = x/R$ and $\eta = y/R$ and
using normalized distance $d = D/R$, we obtain
\begin{equation}  \label{eq:bright}
I = I_0 + a_0 R^{-1/2} J(d) ,
\end{equation}
where the function $J(d)$ is
\begin{equation}
J(d) = \left[\int_{-\infty}^{+\infty}d\eta \int_{0}^{+\infty} d\xi
P(\xi,\eta) \xi^{-1/2} \right] \Big/
\left[\int_{-\infty}^{+\infty} d\eta \int_{-\infty}^{+\infty} d\xi
P(\xi,\eta) \right]
\end{equation}
or, taking into account the symmetry in $\eta$ and the total
intensity
\begin{equation}
\int P(r)d\bm{r} = \frac{\pi R^2}{p-1} ,
\end{equation}
we obtain:
\begin{equation}
J(d) = 2\frac{p-1}{\pi}\int_0^{\infty} d\xi  \int_0^{\infty}d\eta
\xi^{-\frac{1}{2}} \left[1+(\xi-d)^2+\eta^2\right]^{-p} .
\end{equation}
Taking the internal integral over $\eta$ yields
\begin{equation}
J(d)=\frac{p-1}{\pi}\int_0^{\infty}d\xi B\left(\frac{1}{2},
p-\frac{1}{2}\right)\xi^{-1/2}\left[\xi^2-2\xi d +d^2  +1
\right]^{\frac{1}{2}-p} ,
\end{equation}
where $B$ is the beta function. The subsequent integration over
$\xi$ yields
\begin{equation}
J(d)=\frac{\Gamma(p-\frac{1}{2})\Gamma(2p-\frac{3}{2})}{\Gamma(p-1)
\Gamma(2p-1)}(1+d^2)^{\frac{3}{4}-p}{}_2
F_1\left(\frac{1}{2},2p-\frac{3}{2};p;\frac{1}{2}
\left(1+\frac{d}{\sqrt{1+d^2}}\right)\right) .
\end{equation}
Here, $\Gamma$ and ${}_2F_1(a,b;c;z)$ are the gamma function and
the Gauss hypergeometric function, respectively. Figure 2 shows
the function $J(d)$ for several values of $p$. We clearly see from
the figure that the sharpness of the jump in amplitude increases
with increasing source brightness concentration toward the center
during caustic crossing, tending to an infinite point-source limit
for very large $p$.
\begin{figure}
\epsfig{file=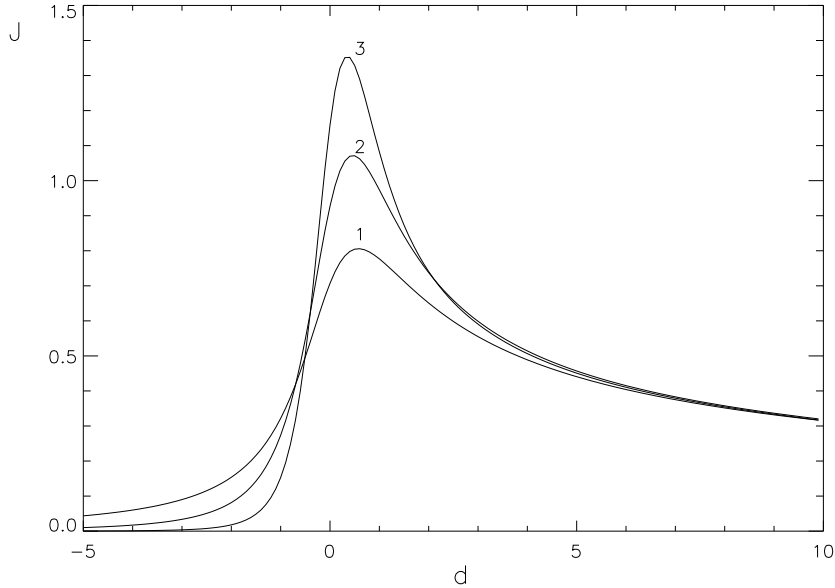,width=11cm}
\caption{Function $J(d)$ for
several exponents $p$: curve $1, 2,$ and $3$ for $p$ = 1.5, 2.0,
and 3.0, respectively}
\end{figure}

For some particular $p$ values, for example, for $p=3/2, 2$ and
$3$, the hypergeometric function can be expressed as a combination
of elementary and other special functions. Thus, we have for
$p=3/2$
\begin{equation}
J(d)_{|p=3/2} = \left[2(1+d^2)((1+d^2)^{1/2}-d)\right]^{-1/2} .
\end{equation}

\section{VARIABILITY ANALYSIS}
We see from Fig. 1 that the intensities of all four quasar images
have varied during the last four years. The largest variations
were observed in image C, which passed its maximum in mid-1999,
and in image A, which reached its maximum in late 1999. The
approach to interpreting the variability of these two images is
the same. Let us assume that, in both cases, the source crosses
the caustic; the brightness distribution must follow the law
(\ref{eq:bright}). The general form of the curve depends on five
parameters:

\begin{enumerate}
\item Contribution $I_0$ from the remaining microimages, which is
assumed to be approximately constant;
\item Caustic strength $a_0$;
\item The time it takes for the source to cross its
radius $\Delta t$ of the source, which is proportional to the
source size $R$;
\item The time $t_0$ of caustic crossing by the source center;
\item Exponent $p$ in the brightness distribution (\ref{eq:P}).
\end{enumerate}

Estimating the five parameters reduces to minimizing the sum of
the squared differences between model and observed light curves
\begin{equation}
\chi^2=\sum_{i=1}^{N}\frac{1}{\sigma_i^2}
\left[I_{model}(I_0,a_0,\Delta t,t_0,p) - I_{obs}\right]^2 .
\end{equation}
The summation is performed over all $N$ points of observations
with weights that are inversely proportional to the squares of the
observational errors $\sigma_i$. The best set of parameters is
sought by the Marquardt method (see, e.g., Chapter 15.5 in the
book by Press et al. 1992).

\subsection{Image A}
The model parameters estimated from the light curve of image A
with 182 data points during the entire observing period 1997-1999
are given in the Table 1. The combination $a_0/\Delta t$ is more
convenient to calculate than the caustic strength $a_0$. The
listed formal accuracies of the parameter estimates should be
considered only as their lower limits.

\begin{table}
\begin{tabular}{l|ccccc}
\hline
Image & $I_0$ & $a_0/\Delta t$ & $\Delta t$ & $t_0$  & $p$\\
\hline
A 1997-1999 & 0.44$\pm$0.01  & 0.73$\pm$0.07  & 91$\pm$4 &
1462$\pm$1 & 1.24$\pm$0.03\\
C 1998-1998 & 0.22$\pm$0.01 & 0.95$\pm$0.15  & -29$\pm$2 &
1390$\pm$1 & 1.10$\pm$0.02\\
\hline
\end{tabular}
\caption{Best-fit model parameters}
\end{table}

Figure 3 shows the model light curve together with measured
values. The 1997 observations are poorly fitted by a single curve,
implying that approaching the caustic did not show up in the first
year. Such a behavior is characteristic of numerical microlensing
models for Q2237+0305, in which one amplification event is often
superimposed on another to form complex light curves. Excluding
the 1997 data from our analysis, while significantly improving the
total $\chi^2$ residual, affects the parameter estimates only
slightly.
\begin{figure}
\epsfig{file=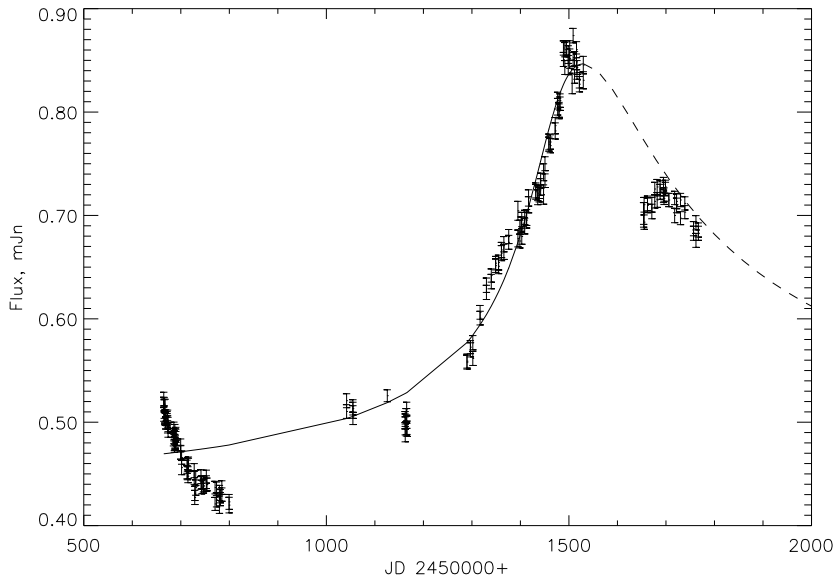,width=11cm}
\caption{Variability of image A
during 1997--2000. Observational data, model, and forecast}
\end{figure}

When the caustic is crossed in image A, an additional pair of
images emerges. A characteristic feature of this direction of
motion is a steep rise in light amplification followed by a
gentler decline. The dashed line in Fig. 3 represents the expected
behavior of image A in the immediate future. We assume that the
brightness will continue to decline and (if no additional causes
of amplification arise) will reach the original 1997 level in two
to three years. The latest observations for 2000 are added for
comparison.

\subsection{Image C}
Attempts to fit the light curve of image C over the entire
observed period have failed. At the same time, excluding 80 data
points for 1997 from our analysis results in quite reasonable
estimates, which are given in the table and shown in Fig. 4.
Interestingly, the best solution corresponds to caustic crossing
in the negative direction, with the pair of images disappearing.
The local minimum of $\chi^2$ corresponding to the motion in the
positive direction is several times greater than the absolute
minimum reached during the motion in the negative direction.

\begin{figure}
\epsfig{file=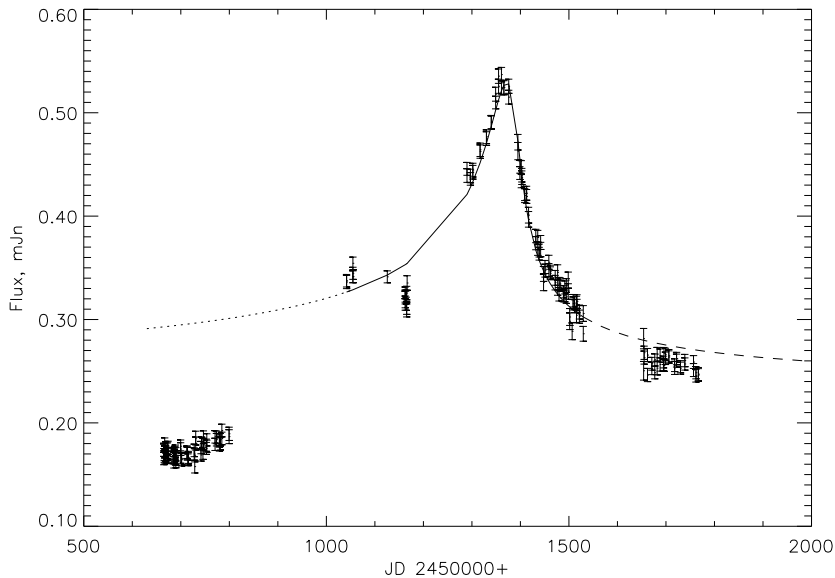,width=11cm} \caption{Light curve of image C
during 1997--2000. Observations, model (dotted line), and
forecast}
\end{figure}
The inability to fit the entire observed period by a single curve
becomes more understandable in light of the recent results by
Wyithe et al. 2000. These authors argue that the intensity
variations in image C have been composite in pattern during the
last four years, and, what is quite possible, another caustic
crossing was overlooked during the observing seasons in 1997 and
1998.

The dotted and dashed lines indicate the best extrapolation
computed in the model with single caustic crossing in 1997 and our
forecast until the end of 2000, respectively. No appreciable
intensity variations in image C are expected in the immediate
future.

\section{THE SIZE AND STRUCTURE OF \\
THE QUASAR EMITTING REGION} For the quasar radius estimates
$\Delta t$ in time units to be transformed to linear sizes $R$, we
must know the apparent quasar velocity; only the velocity
component $v_\perp$ is perpendicular to the caustic line is
important. Of course, the exact velocity is unknown. Nevertheless,
a statistical analysis of the time derivatives of brightness
variations by Wyithe et al. 1999 yielded an upper limit of $v <
500$ {\it km/s}. The perpendicular velocity component can only be
lower than this value. The most probable value of $v_\perp$ was
found to be $300$ {\it km/s}. Given the angular distances from the
observer to the quasar $D_{os}$ and to the lens $D_{ol}$ in the
model of a flat Universe with $H_0 = 60$ {\it km/s} $Mpc^{-1}$,
the most probable source radius is estimated to be
\begin{equation}
R = v_\perp \Delta t \frac{D_{os}}{D_{ol}} \sim 2.0 \cdot 10^{13}
\left(\frac{v_\perp}{300 \textit{km/s}}\right)\left(\frac{\Delta
t}{1 \textit{day}}\right) \textit{cm} .
\end{equation}
Using the crossing time $\Delta t_A = 90$ {\it days} for image A
gives the most probable quasar radius $R = 1.8 \cdot 10^{15}$ {\it
cm}, while substituting the crossing time for image C, $\Delta t_C
= 30$ {\it days}, reduces this estimate to $R = 6 \cdot 10^{14}$
{\it cm}. At the same time, the velocity constraint $v < 500$ {\it
km/s} together with the crossing time $\Delta t_C$ give an upper
limit on the source radius, $R \leq 10^{15}$ {\it cm}.

Another parameter that characterizes the mass distribution is the
exponent $p$. It follows from the table that its computed value
lies between 1.1 and 1.25. It is interesting to note that
different models for the quasar structure can lead to different
dependences of the emissivity on distance from the center. For
instance, a power-law dependence with $p \leq 0.5$ follows from
the model of an optically thin accretion disk (Manmoto et al.
1997). At the same time, the standard model of an optically thick
accretion disk yields an $r^{-3}$ dependence (Shakura and Sunyaev
1973), which changes to $(r^2+R^2)^{-3/2}$ for a finite radius.
Our estimate $p \sim 1.2$ favors the standard model, but more
complex accretion-disk models should be analyzed to achieve better
agreement.

\section{DISCUSSION}
Let us consider the legitimacy of some assumptions made here. The
main assumption is associated with the hypothesis of caustic
crossing. Numerical calculations show that there are two effects
capable of causing a significant increase in amplification. Apart
from fold-caustic crossing, the source can also pass near the
caustic cusp. However, events of the second type for the images of
Q2237+0305 are several times less probable than those of the first
type (Wambsganss et al. 1992; Lewis and Irwin 1996) and, in
general, are more symmetric. These two properties can serve as a
statistical justification for using the hypothesis of caustic
crossing.

A power-law model with a finite core radius is used to calculate
the quasar size and structure. Three models of a caustic-crossing
source are encountered in the literature: a homogeneous disk, a
Gaussian source (Schneider and Weiss 1987), and a
$(1-r^2/R^2)^{1/2}$ distribution (Schneider and Wagoner 1987). All
of them are completely determined by their radii. The
$(1+r^2/R^2)^{-p}$ model differs radically in that it is a
two-parameter model. Estimating the rate of brightness decline $p$
allows us to choose between different models for the quasar
structure.

Allowance for the spatial orientation of the emitting region
appreciably complicates the analysis. Two additional parameters
associated with the orientation-angle components of the initial
circular source appear. The first and second parameters determine,
respectively, the apparent-ellipse eccentricity and the angle of
motion with respect to the perpendicular to the caustic (in this
case, allowance for the direction of motion does not reduce to a
simple substitution of the transverse velocity). Besides,
accretion-disk rotation can cause an additional asymmetry of the
emitting region in the spectral range considered due to the
Doppler effect.

In view of many influencing factors, the possibility of
reconstructing, at least in principle, the one-dimensional quasar
profile (along the $x$-axis) as the solution of an integral
equation (Grieger et al. 1991; Mineshige and Yonehara 1999; Agol
and Krolik 1999) seems of great interest. However, this is a
separate, independent problem, which is yet to be applied to an
actual monitoring.

The simple amplification behavior $\sim x^{-1/2}$ for a point
source located in the caustic region is possible only when several
conditions are satisfied. The radius of caustic curvature must be
considerably larger than the source size (see Fluke and Webster
1999 for curvature allowance). The caustic must be isolated lest
the source cover several caustics simultaneously. In addition, it
is implied that there is no large-scale time gradient in
amplification variations. Thus, for example, introducing a
constant slope as an additional free parameter allows a model
curve of image C to be easily fitted to the entire observing
period 1997--1999. However, since there were no such large
gradients throughout the entire 15-year-long monitoring history of
Q2237+0305 and since their physical origin is not completely
understood, we have to abandon the additional parameter.

Finally, the problem can be further complicated by intrinsic
quasar variability. In general, the latter is rather difficult to
separate from microlensing variability. However, given that the
delays between the images in our cases do not exceed several days,
intrinsic quasar variability must be repeated in all quasar images
(with individual amplification factors) virtually simultaneously.
The fact that the brightness variations in all four components are
not synchronous provides circumstantial evidence that intrinsic
quasar variability is negligible in this source.

Despite the possible complicating factors listed above, we have
every reason to believe that the simple model used here is capable
of faithfully reproducing the observational data, and that its
implications deserve a careful analysis.

After this paper was mainly complete, Wyithe et al. 2000
independently published a preprint where they also interpreted the
OGLE-group observations of Q2237+0305. These authors analyzed the
light curves by using statistical methods based on conditional
probability distributions.

Wyithe et al. 2000 focused mainly on computing the probability of
occurrence of brightness bursts with observed parameters and on
estimating the possibility of subsequent bursts. The conclusion
that there is an additional overlooked brightness burst in image C
associated with caustic crossing between the observing seasons of
1997 and 1998 seems of considerable interest. Such an event can
account for the difficulties of modeling the light curve for image
C in terms of the model with a single caustic crossing. Wiythe et
al. 2000 also expect another caustic crossing in image C 500 days
after the 1999 summer maximum (with a large uncertainty of $\sim
100 - 2000$ days, though).

The amplification in image A in late 1999 is interpreted as
caustic crossing in the negative direction, just as we did here.
However, the peak of image C in mid-1999 is assumed to be caused
by the passage of the source near the caustic cusp. The choice
between the two interpretations of image C variability could be
made by analyzing color variations of the source, which are much
larger during caustic crossing than during cusp passage. A color
analysis could be performed by invoking additional data of the
simultaneous monitorings at the Apache Point and Maidanak
Observatories (Bliokh et al. 1999) through various filters during
the same observing period.
\\
\\
{\it Acknowledgments.} I wish to thank the OGLE group for
organizing and carrying out the monitoring of Q2237+0305 and for
providing access to the data before their publication.

\section*{REFERENCES}

Agol E. and Krolik J., Astrophys. J. {\bf 524}, 49 (1999)\\
Blandford R. D. and Narayan R., Astrophys. J. {\bf 310}, 568
(1986)\\
Bliokh P. V., Dudinov V. N., Vakulik V. G.  et al., Kinematika
Fiz. Nebesnykh Tel {\bf 15}, 338 (1999)\\
Chang K. and Refsdal S., Astron. Astrophys. {\bf 132}, 168
(1984)\\
Corrigan R. T., Irwin M. J., Arnaud J. et al., Astron. J. {\bf
102}, 34 (1991)\\
Fluke C. J. and Webster R. L., Mon. Not. Roy. Astron. Soc. {\bf
302}, 68 (1999)\\
Grieger B., Kayser R., and Schramm T., Astron. Astrophys. {\bf
252}, 508 (1991)\\
Irwin M. J., Webster R. L., Hewett P. C. et al., Astron. J. {\bf
98}, 1989 (1989)\\
Lewis G. F. and Irwin M. J., Mon. Not. Roy. Astron. Soc. {\bf
283}, 225 (1996)\\
Manmoto T., Mineshige S., and Kusunose M., Astrophys. J. {\bf
489}, 791 (1997)\\
Mineshige S. and Yonehara A., Publ. Astron. Soc. Jpn. {\bf 51},
497 (1999)\\
\O stensen R., Refsdal S., Stabell R. et al., Astron. Astrophys.
{\bf 309}, 59 (1996)\\
Press W, Teukolsky S., Vetterling W. et al. Numerical Recipes in
FORTRAN (Cambridge Univ. Press, Cambridge, 1992, 2nd ed.), p.
675\\
Schneider P. and Wagoner R. V., Astrophys. J. {\bf 314}, 154
(1987)\\
Schneider P. and Weiss A., Astron. Astrophys. {\bf 171}, 49
(1987)\\
Shakura N. I. and Sunyaev R. A., Astron. Astrophys. {\bf 24}, 337
(1973)\\
Wambsganss J., Witt H. J., and Schneider P., Astron. Astrophys.
{\bf 258}, 591 (1992)\\
Wozniak P. R., Alard C., Udalski A. et al., Astrophys. J. {\bf
529}, 88 (2000)\\
Wyithe J. S. B., Webster R. L., and Turner E. L. , Mon. Not. Roy.
Astron. Soc. {\bf 309}, 261 (1999)\\
Wyithe J. S. B., Turner E. L., and Webster R. L. ,
astro-ph/0001307 (2000)\\
Zakharov A. F., Gravitational Lenses and Microlenses (Yanus,
Moscow, 1997)\\
Zakharov  A. F. and Sazhin M. V., Usp. Fiz. Nauk {\bf 168}, 1041
(1998) (Phys. Usp. {\bf 41}, 945 (1998))\\
\end{document}